# A new approach to measure the scientific strengths of territories[1]

**Giovanni Abramo***
*Laboratory for Studies of Research and Technology Transfer. Institute for System Analysis and Computer Science - National Research Council of Italy (IASI-CNR) Viale Manzoni 30, 00185 Rome, Italy - giovanni.abramo@uniroma2.it*

**Ciriaco Andrea D'Angelo**
*University of Rome "Tor Vergata", Dept of Engineering and Management Via del Politecnico 1, 00133 Rome, Italy - dangelo@dii.uniroma2.it*

**Flavia Di Costa**
*University of Rome "Sapienza", Dept of Computer, Control and Management Engineering. Via Ariosto 25, 00185 Rome, Italy - dicosta@dis.uniroma1.it*

**Abstract**

The current work applies a methodology for mapping the supply of new knowledge from public research organizations, in this case from Italian institutions at the level of regions and provinces (NUTS2 and NUTS3). Through the analysis of scientific production indexed in the Web of Science for the period 2006-2010, the new knowledge is classified in subject categories and mapped according to an algorithm for the reconciliation of authors' affiliations. Unlike other studies in the literature based on simple counting of publications, the present study adopts an indicator, Scientific Strength, which takes account of both the quantity of scientific production and its impact on the advancement of knowledge. The differences in the results that arise from the two approaches are examined. The results of works of this kind can inform public research policies, at national and local levels, as well as the localization strategies of research-based companies.

---

# 1. Introduction

Since the early 1990s, the concept of regional innovation systems has attracted the attention of scholars and policy makers as a possibly ideal framework in support of comprehending innovation processes in regional economies. There is no unequivocal definition of the concept of the regional innovation system: according to the general understanding it entails a set of interacting private and public interests, institutions and organizations that operate for the generation, dissemination and exploitation of knowledge (Doloreux, 2005; Cooke, 2004). The origin of the concept is traced to two different theoretical spheres: systems innovation and regional science. From regional science, it takes the concepts of the role of geographic proximity (benefits deriving from advantages in localization and spatial concentration) and a set of territorial rules, conventions and norms governing the rise of processes of creation and dissemination of knowledge (Kirat and Lung, 1999).

One of the most important applications that derive from the overall theoretical approach is the analysis of interactions between actors in the process and in particular the relations between the industrial and the research spheres, including their potential mediation by organizations devoted to technological transfer. Another important actor, the government, has been envisaged in the so called "triple helix" model of innovation (Leydesdorff and Fritsch, 2006; Leydesdorff and Etzkowitz, 1996). The triple helix model "is not first specified in terms of domains (e.g., national systems) or specific functions (e.g., knowledge production), but allows for interaction effects among domains and specific synergies among functions and institutions" (Leydesdorff, 2010).

The literature on the theme has certainly given a strong boost to the so-called endogenous approach to local and regional development policy, based on the idea that regional development and the resulting economic growth must be driven by endogenous forces in the form of "a highly educated workforce and knowledge and technologies developed in the region" (Todtling, 2010). At the regional level, universities are considered as the core knowledge-producing bodies, capable of a primary role in activating the innovation and development agenda, through their placement as key elements in innovation systems and principle providers of knowledge for the industrial sector (Kitagawa, 2004; Thanki, 1999; Garlick, 1998; Foray and Lundvall, 1996). The companies of a given territory will use the knowledge produced in universities in different manners. The smallest ones can benefit from the spillovers of university knowledge, since they have fewer resources for active R&D to produce such knowledge on their own (Acs et al., 1994). Regional high-technology firms also benefit from university knowledge (Audretsch et al., 2005), as demonstrated by research detecting that, on a regional basis, there is a significant correlation between concentrations of high-technology industries and university research in high-technology fields (Nagle, 2007).

The region thus becomes the promoter of its own development and is seen as the territorial level of reference to engage growth, through local knowledge spillovers, intra-regional networks and labor mobility (Martin and Sunley 1998; Krugman 1991). Such lines of reasoning have clearly been influential, as seen in the fact that over the past 10 years, European supranational policies regarding innovation have focused strongly on the local dimension. These policy trends are bolstered by the empirical evidence and studies demonstrating that the region is the crucial level for knowledge creation and diffusion, learning and innovation.

In this context, one of the themes of certain interest for scholars in the subject is the analysis of the contributions of public research organizations and universities, which face increasing demands to integrate their traditional teaching and research activities with a third mission: support for competiveness of the local industries, through licensing of inventions, spin off creation, research collaborations and partnership with private companies, etc. (Glasson, 2003; Thanki, 1999). In particular, Maier and Luger (1995) identify eight activity areas that have essential benefits for economic development: i) creation of knowledge, ii) human capital creation, iii) transfer of existing know-how, iv) technological innovation, v) capital investment, vi) regional leadership, vii) influence on regional milieu, and viii) knowledge infrastructure production. However it should be noted that on this subject, scholars in the field are not always in agreement. Feller (2004), for example, holds that beyond the tangible and easily measurable impacts ensured by activities in technological transfer, the true contributions of universities to economic growth lie in the creation of public knowledge and in the education and training of a qualified work force. Others, including Stoneman and Diederen (1994), focus their attention on knowledge diffusion and the importance of adopting effective means for transferring knowledge to both private and public sectors, through both formal and informal dissemination mechanisms.

The localization of public research institutions within a particular nation has historic, economic, and sociological origins, and more recently is ever more influenced by policy and strategic decisions. Whatever the origin of the current territorial distribution of new knowledge suppliers, the policy maker certainly has interests in monitoring evolution in the mapping of scientific production, for purposes of understanding and decision-making regarding the distribution and maximization of its benefits. Similarly, for research-based companies in the private sector, the territorial mapping of new knowledge can inform efficient choices in the localization of R&D activities.

But is it possible to measure new knowledge creation and diffusion at the level of restricted geographic areas? In the literature, the studies that have attempted this task have typically adopted a bibliometric approach based on analysis of the geographic distribution of scientific production, as indexed in the major bibliometric databases. This approach assumes that scientific publication in international journals is the principal form of dissemination of results from research activity, as conducted by universities and research organizations in general. Frenken et al. (2009) offer a particularly useful review of the full range of scientometric studies analyzing the spatial dimension of scientific production, beginning from the pioneering works by Narin and Carpenter (1975) and Frame et al. (1977). This latter work, under the suggestive title "The distribution of world science", is based on data from the ISI Science Citation Index on the distribution of output from 117 countries and in 92 disciplines, over one year (1973). More recent studies, employing similar methodologies, have primarily concerned the spatial concentration of scientific production, which seems to have remained high for the industrialized nations of the OECD: these nations thus continue to account for the major share of world output (May, 1997; Adams, 1998; Cole and Phelan, 1999; Glänzel et al., 2002; King, 2004; Horta and Veloso, 2007), despite a rapid increase in scientific production from China (Leydesdorff and Zhou, 2005). Analyses at the regional level have been less frequent: one case is the work by Matthiessen and Schwarz (1999), on the analysis of aggregated publication records for European metropolitan areas, for the years 1994-1996.

Some scholars have also proposed analyses based on the spatial distribution of highly-cited publications, primarily for the identification of centers of excellence at the regional level (Bonitz et al., 1997; Batty, 2003). More recently, a work by Bornmann and Leydesdorff (2011), based on the Web of Science (WoS) data, identifies cities where top-10% most-highly-cited papers were published more frequently than would be expected, offering visualization of the results via Google Maps. In very similar manner, Bornmann et al. (2011) present methods for mapping centers of excellence around the world, in this case using Scopus data. Excellence in single scientific fields is identified, revealing agglomerations in regions and cities where highly-cited papers (top-1%) were published. Shifting the focus from cities to regions, Bornmann and Waltman (2011) use visualization methods (density maps) to detect regions of excellence at the global level, focusing on the top 1% of 2007 papers indexed in Scopus. Very recently, Bornmann et al. (in press) presented a web application to identify research centers of excellence by field worldwide, using publication and citation data.

In this work we propose to advance the literature on the theme, with a study that returns to the analysis of the spatial distribution of research activity, but with a new indicator, based on the standardized citations received by the scientific portfolio of research institutions for a territory. Differently from previous studies, which count publications or highly-cited ones, we compare the scientific strength of territories in each scientific field by the overall standardized citations. We apply the proposed methodology to the case of Italy, trying to answer the following research questions:

- How is the production of scientific research distributed across national territory?
- Which regions/provinces lead in scientific production by field?
- Is it possible to detect geographic concentrations of activity in specific fields?
- Would there be different results if spatial analysis were conducted through simple

counting of publications, rather than from citations?

The study of the territorial distribution of public supply of knowledge can certainly orient the necessary actions of decision makers responsible for this area, at the levels of nation, region and province. In this sense, the authors hold that the proposed methodology can in fact provide the answers to the above research questions, and that it thus represents a useful contribution, going beyond the specific national case where the new approach is applied.

The study is organized as follows: the next section presents the methodology for the analysis (field of observation, sources and dataset). Section 3 presents and discusses the results of the analyses, while Section 4 concludes with a summary of the significant findings and the authors' comments.

## 2. Methodology

In regards to the territorial framework for analysis, we refer to the Nomenclature of Territorial Units for Statistics (NUTS)[2]. In Italy the aggregations provided under legislation for the national units of political and administrative decentralization are the Regions (NUTS 2) and Provinces (NUTS 3). The scientific production of public research institutions is extracted from the Italian Observatory of Public Research (ORP)[3], a database developed and maintained by the authors and derived under license from the Thomson Reuters WoS. Beginning from the raw data of the WoS, and applying a complex algorithm for reconciliation of bibliometric addresses, each publication is attributed to the organizations of its co-authors, and consequently to the territory where they work. The algorithm is based on a controlled vocabulary of over 30,000 rules (D'Angelo et al., 2011)[4]. Unlike the arts and humanities and some fields of the social sciences, in the hard sciences the prevalent form of codification for research outputs is publication in scientific journals. Other forms of output are often followed by publications that describe their content in the scientific arena. Thus analysis of publications alone permits derivation of mapping that is certainly representative of the new knowledge produced by public research organizations, providing that the field of observation is limited to the subject categories of the hard sciences[5] (a total of 167 categories, according to WoS classification[6]).

---

[2] NUTS is a geocode standard for referencing the subdivisions of countries for statistical purposes. The standard is developed and regulated by the European Union, and thus only covers the member states of the EU in detail.

[3] www.orp.researchvalue.it. Last accessed on January 14, 2014.

The data extracted thus concern the scientific production achieved in the given subject categories over the 2006-2010 period, by all national public research organizations, meaning all Italian universities (95), research institutions (76) and research hospitals (200). This dataset of 2006-2010 Italian scientific production (articles, reviews, proceeding papers, letters) in the hard sciences consists of roughly 260,000 publications.

To evaluate the public supply of knowledge we consider not the simple counting of publications produced, but rather their real value in terms of impact on the advancement of knowledge. As proxy of value, bibliometricians adopt the number of citations received by the publication. In particular, we use a relative indicator, named Article

---

[4] As an example, the rules resolve 142 variants of "University of Rome 'Tor Vergata'", detected in WoS bibliometric affiliations for the period under examination.
[5] Biology, biomedical research, chemistry, clinical medicine, earth and space sciences, engineering, mathematics, physics.
[6] http://ip-science.thomsonreuters.com/cgi-bin/jrnlst/jlsubcatg.cgi?PC=K, last accessed on January 14, 2014.

Impact Index (AII), given by the ratio between the number of citations received by a publication (as of 31/12/2011) and the average of the citations for all the other national publications of the same year and subject category[7] (Abramo et al., 2012). For each subject category, the values of AII are successively aggregated at the Provincial level (NUTS3) and the next higher level of the region (NUTS2), to obtain an indicator named Scientific Strength (SS) given by the sum of the Impact Index (AII) of all the publications produced in the particular territory. The publications co-authored by scientists working in organizations of the same territory are counted only once for that territory. In assigning a publication to a territory we do not adopt fractional counting in function of the number of authors. The reasoning for these last-described procedures is that a publication represents new knowledge produced in a territory independently of the number of people in that territory that contributed to its production. For publications in multi-category journals, to each subject category is attributed a fractional value of AII, equal to the inverse of the number of subject categories included in the journal.

**3. Results and analysis**

**3.1 Distribution of the supply at the level of regions**

In terms of administrative structure, the Italian state is subdivided in 20 regions and 110 provinces (Table 1). Four regions (Valle d'Aosta, Umbria, Basilicata, Molise) have less than a million habitants out of the total population of 61 million, while at the next level there are 61 provinces with less than 40,000 inhabitants and only 10 with more than a million.

---

[7] The subject category of a publication corresponds to that of the journal where it is published. For publications in multidisciplinary journals the scaling factor is the average of the scaling factors of each subject category of the journal.

[Insert Table 1 here]

In our first analysis of the public supply of new knowledge on a geographic basis we examine, for each region, the prevalent subject categories. For this, alongside each region name, Table 2 presents the first three subject categories by incidence of SS in the national total. As we would expect, in the small regions the supply of new knowledge is modest with respect to the national total, even in the prevalent subject categories: in Umbria, Basilicata and Molise, research in the prevalent subject category (Engineering, petroleum for all three regions) does not exceed 8% of national SS. Research in the prevalent category in Valle d'Aosta (Operations research & management science) produces SS that does not reach 1% of the national total. In contrast, the large regions are the seat of very concentrated research activity in the prevailing categories: Lombardy alone produces almost half of national SS in Ornithology (47.1%); Tuscany produces 38.5% of SS in Andrology; Lazio achieves 37.5% of that in Tropical Medicine.

[Insert Table 2 here]

The data concerning concentration of research activity in certain subject categories, and in certain regions (usually large ones), stimulates the deeper analysis presented in Table 3. The table lists the first 20 subject categories for regional concentration of SS, ordered by decreasing cumulative value referring to the first three regions. The highest value of cumulative incidence is for Ornithology, a field where more than three quarters of the advancement in knowledge (77.8%) is achieved by Lombardy, Emilia Romagna and Sicily. The values of Gini coefficient[8] (shown in the last column of Table 3; all in the interval 0.6-0.8) demonstrate the strong territorial concentration of research activity in these 20 subject categories, which concern five disciplines. Specifically, a full 10 subject categories out of 20 fall under Clinical medicine, five under Engineering, two each under Biology and Biomedical research and one under Physics.

[Insert Table 3 here]

From the data in columns 3, 4 and 5 in Table 3, we observe that the 60 positions referring to prevalent subjects are occupied by 11 regions out of the 20 possible: six are situated in the north (three each in the northwest and northeast), two in central Italy and three in the south. In detail, in Column 3, the highest frequency is registered for Lombardy (prevailing in nine subject categories, the majority in the Life sciences disciplines), followed by the three subject categories for Lazio (Clinical medicine) and Emilia Romagna (two in Clinical medicine and one in Engineering). Similarly, the data in Column 4 again show Lazio and Lombardy as regions prevailing in seven out of the 20 subject categories listed.

Table 4 is structured like the preceding one but refers to the 20 subject categories with SS least concentrated at the level of regional distribution. The values of cumulative incidence in the first three regions are less than 40% in all cases, dropping to a minimum of 3.2% for Marine & freshwater biology, while the values of Gini coefficient vary in the interval 0.4-0.5. The 20 subject categories indicated belong to five disciplines: eight are in Engineering, six in Biology, two each in Chemistry, Earth and space sciences and Physics. It is also notable that there is a total absence of subject categories of Clinical medicine and Biomedical research. Observing the regions indicated in columns 3, 4 and 5 of the table, we note that the 60 positions referring to the SS least concentrated subjects categories are occupied by 12 regions, generally large ones (as we would expect): six are in the north (three each in the northwest and northeast), two in central Italy and four in the south. Again in this case, the highest frequency is for Lombardy, followed by Emilia Romagna.

---

[8] Gini coefficient is the most commonly used measure of inequality. It varies between 0, which reflects complete equality and 1, which indicates complete inequality (one entity has all the measure, all others have none).

[Insert Table 4 here]

**3.2 Distribution of the supply at the level of provinces**

We now repeat the type of analysis conducted for the regions, but at the level of provinces. Table 5 presents the 20 subject categories with SS most concentrated at the level of Provincial distribution, in order of decreasing value of cumulative incidence of SS for the three leading provinces.

We note that the values of cumulative incidence vary from a maximum of 66.0% for Engineering, marine to a minimum of 44.0% per Health care sciences and services; the values of Gini coefficient are all greater than 0.8.

In general, the subject categories indicated seem very specialized, of "niche" type, or concentrated in particular territories because of the contextual presence of specific public research agencies, clinical research institutes or universities.

In particular, the data for Engineering, marine show that for this subject category, two thirds of knowledge advancement is produced from research activity conducted in only three provinces (Trieste, Pavia, Messina). In terms of geographic concentration, we also see interesting cases for Ornithology, Sport sciences and Tropical medicine. In these three subject categories, a third of national SS is provided by research conducted in a single province: Milan in the first case, Rome in the other two.

In effect, Column 3 indicates that these two provinces are prevalent for national research in many of the subject categories with high concentration of activity, and in particular in six specialties of Clinical medicine for Rome and five of Biology and Biomedical research specialties for Milan. The high frequency of life science subject categories (among the 20 presented in the table) confirms the situation that emerged in the preceding section, concerning the high territorial concentration of research conducted in some specialties of these disciplines.

Finally, from the lists in columns 3, 4 and 5, we observe that the 60 positions referring to prevalent subjects are occupied by 20 provinces belonging to 12 different regions: six are from the north (three each from the northwest and northeast), two from central Italy and four from the south. Of the 20 provinces that appear in the table, only three have less than 400,000 inhabitants (Trieste, Sassari, Ravenna).

Naturally, the territorial concentration of research activity is typical of some subject categories but not all. The maps that follow permit comparison of the Provincial distribution of SS in two distinct subject categories of Biology (Ornithology; Marine and freshwater biology - Figure 1) and two of Engineering (Engineering, marine;

Engineering, civil - Figure 2). These subject categories are those within the same discipline that register the maximum and minimum values of Gini coefficient, meaning the maximum and minimum territorial concentration of supply of new knowledge, as is clear from the maps.

[Insert Table 5 here]

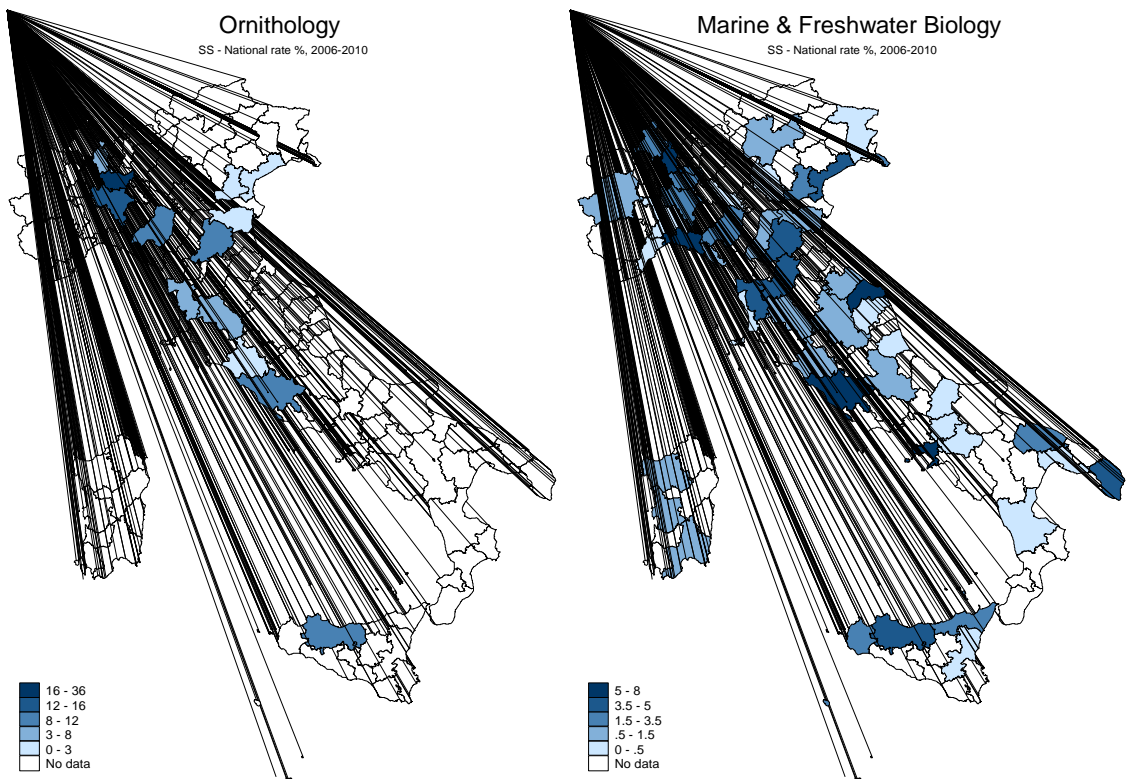

*Figure 1: Provincial distribution of SS in two subject categories of Biology*

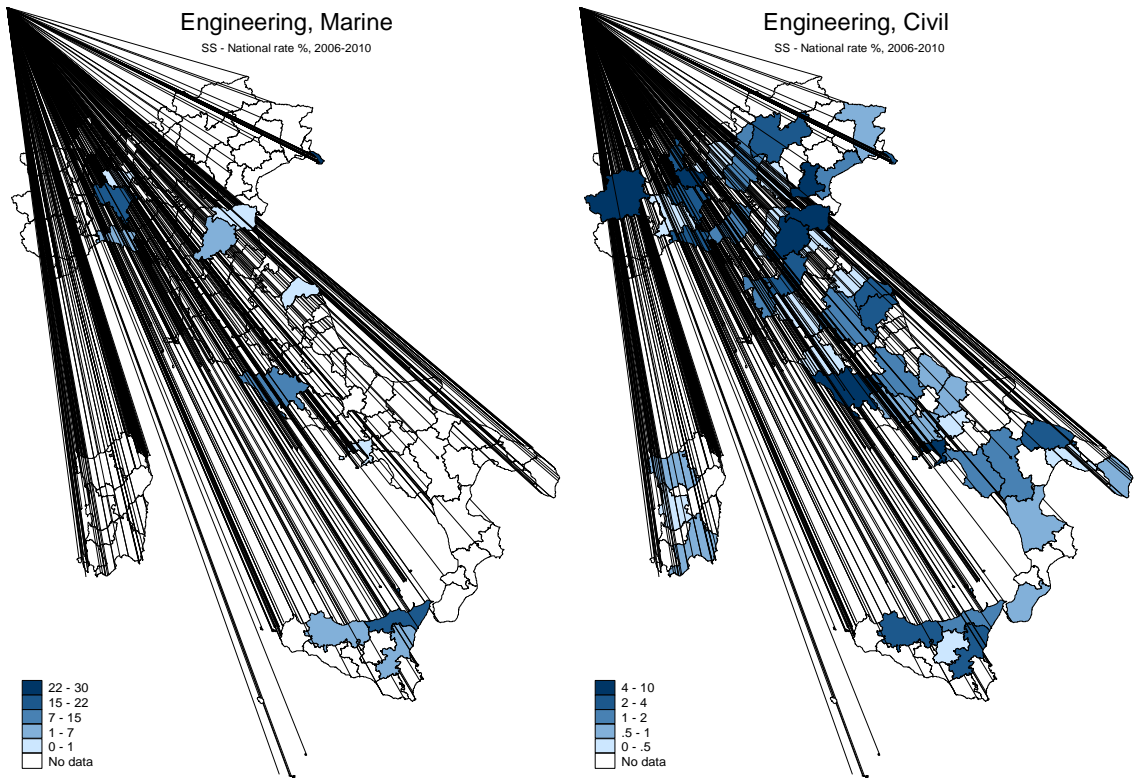

*Figure 2: Provincial distribution of SS in two subject categories of Engineering*

### 3.3 Distribution of supply of new knowledge standardized by socio-economic data

The analyses conducted to this point obviously reflect certain structural characteristics of the territories analyzed. In particular, it very clearly emerges that the distribution of research activity is consistent with the socio-economic structure of the territory: the largest regions and provinces are those where the production of new scientific knowledge tends to concentrate, given the presence of a larger mass of organizations, infrastructure and assets dedicated to research.

In this section we repeat the preceding analyses, but now standardizing the values of SS to the territorial population, thus obtaining a value for new knowledge produced per inhabitant. The data thus obtained are then related to the national average, to demonstrate certain characteristics of the territorial distribution of the public supply of new knowledge. Table 6 shows the 20 Subject category-Region pairings with the maximum level of difference of SS per inhabitant compared to the national average. We observe that for these pairings, the SS per inhabitant is at least five times the national average.

In these same 20 parings, certain regions repeat: this is the case for Friuli Venezia Giulia (six times), Basilicata (five) and Liguria, Molise and Trentino-Alto Adige (twice each). We further observe that these are small regions, both for territorial dimension and number of inhabitants. Similarly, there is a repetition of subject categories: Engineering, petroleum is present in three separate pairings, with Molise, Basilicata and Umbria, while Forestry repeats with Molise and Basilicata. The subject categories present in the 20 pairings belong primarily to the disciplines of Engineering (7) and Physics (5).

[Insert Table 6 here]

We now ask what variations are introduced into the analyses of concentration of new knowledge if we consider simple counting of the publications rather than citations. To respond, we repeat the analysis just presented concerning the output per inhabitant (Table 7). The comparison between the data of Table 6 and Table 7 reveals a certain level of correlation between the results of the two analyses.

However, some substantial differences also emerge. Four of the 17 subject categories listed in Table 6 do not appear in Table 7: this occurs for Remote sensing; Imaging science & photographic technology; Engineering, geological; and Computer science, cybernetics (all belonging to the disciplines of Physics or Engineering). For other three subject categories (Engineering, marine; Entomology; Engineering, petroleum), the region that appears in the two ranking lists is different. Finally, of the eight regions listed in Table 6, one (Umbria) does not appear among those listed in Table 7. In fact, the analysis conducted with the indicator SS per inhabitant lists Engineering, petroleum in Umbria, but this entire pairing is missing in the analysis based on output per inhabitant, substituted by Dentistry, oral surgery & medicine in Abruzzo. Thus it is evident that the territorial analysis of scientific output based on simple counting of the publications produced by organizations of a given region gives mapping that is markedly different than that which can be obtained using citations as a proxy of the real advancement of knowledge in a given discipline.

[Insert Table 7 here]

We conclude the analysis at the regional level, now standardizing SS to regional GDP (average value, 2006-2010). The SS/GDP ratio for each region is related to the national average and we extract the 20 Subject Category-Region pairs that present the maximum difference from the national reference (Table 8). Compared to the data in Table 6, we observe ratios to the national average that are higher, ranging from a minimum of 713.1% (Engineering, petroleum in Umbria) and a maximum of 2,267.9% (again for Engineering, petroleum, but in Molise), compared to the range of 639% to 1,753.1% seen in Table 6.

In the 20 pairings listed in Table 8, some regions are repeated: this occurs for Basilicata (six times), Molise (five), and Friuli Venezia Giulia (four), which are all small regions. The data in Table 8 very clearly indicate a situation readily superimposed on the preceding analysis, with 13 of the 16 subject categories listed also present in Table 6.

[Insert Table 8 here]

The analysis of national distribution of Scientific Strength per inhabitant can also be repeated at the Provincial level[9]. Again comparing the data to the relevant national average, we see a number of cases that are clearly interesting. Table 9 shows the 20 Subject Category-Province pairs with the maximum level of difference of SS per inhabitant compared to the national average.

[Insert Table 9 here]

We observe that in all cases the SS per inhabitant is much greater than the national average. In the 20 pairings, some provinces repeat: this occurs for Trieste (nine times), Pisa and Siena (three), and also Viterbo (two). These are all small provinces, both for number of inhabitants and other macroeconomic characteristics: Pisa, the most populous, barely exceeds 400,000 inhabitants. Similarly to the previous analyses, we see a repetition of subject categories in the case of Engineering, marine which occurs in two different pairings (Trieste and Pavia). The 20 subject categories identified belong primarily to Engineering (eight cases), Physics (four) and Earth and space sciences (three).

---

[9] Analysis of SS/GDP at the provincial level is not possible, due to the lack of the relevant data on GDP.

## 4. Conclusions

In recent decades there has been a marked intensification of studies on the national distribution of the public supply of knowledge, intended to orient action in this area by decision-makers, at their specific levels of competence: nation, region or province. This intensification also arises from recent developments in bibliometrics, which make it possible to produce detailed and exhaustive mapping of the scientific production achieved by research organizations in a given territory. Given this situation, the authors have presented a mapping system based on the citation data concerning the 2006-2010 scientific production achieved by the totality of Italian universities and research organizations, in an attempt to respond to several research questions of real interest to policy makers.

The analysis demonstrates that there is a significant territorial concentration of research activity in some specific subject categories, particularly in the Life sciences and Engineering. As we would expect, it also emerges that the regions and provinces that contribute more to production of new knowledge are the larger ones. However, standardizing the values of the indicator used to quantify the new knowledge produced with respect to socio-economic characteristics of a region/province (number of inhabitants, GDP), we observe interesting concentrations of activity, above all in the small territories.

One of the obvious areas for deepening the current research concerns the measurement of indices of sectorial specialization, which would permit further characterization of the scientific activity of each territory. The current authors will certainly engage in this research, examining the case of provinces and regions in Italy.

Another notable result of the current work concerns the sensitivity of mapping with respect to the type of bibliometric indicator employed. The analysis of the literature indicates that the prevalent method is that based on the simple counting of publications: the comparison of the results from this type of analysis and that based on standardized citations (applied in this work), reveals the presence of significant differences. It is thus evident that the approach based on simple counting of publications should be substituted by an approach such as that seen in the current work, which takes account of the real impact of publications in the advancement of knowledge in a given disciplinary environment.

In conclusion, the authors trust that the methodology advanced for response to the research questions will go beyond interest for the national case examined, representing a useful contribution to the development of supports to the decision-making contexts of policy makers, acting at various territorial levels of competence.

*Table 1: List of Italian regions and provinces; population data averaged over the years 2006-2010*

| Macro-area | Region | Inhabitants (x 1,000) | Provinces |
|---|---|---|---|
| Northwest | Liguria | 1,612 | Genoa; Imperia; La Spezia; Savona |
| Northwest | Lombardy | 9,646 | Bergamo; Brescia; Como; Cremona; Lecco; Lodi; Mantua; Milan; Monza and Brianza; Pavia; Sondrio; Varese |
| Northwest | Piedmont | 4,395 | Alessandria; Asti; Biella; Cuneo; Novara; Turin; Verbania; Vercelli |
| Northwest | Valle D'Aosta | 126 | Aosta |
| Northeast | Emilia Romagna | 4,284 | Bologna; Ferrara; Forli-Cesena; Modena; Parma; Piacenza; Ravenna; Reggio Emilia; Rimini |
| Northeast | Friuli Venezia Giulia | 1,222 | Gorizia; Pordenone; Trieste; Udine |
| Northeast | Trentino Alto Adige | 1,007 | Bolzano; Trento |
| Northeast | Veneto | 4,828 | Belluno; Padua; Rovigo; Treviso; Venice; Verona; Vicenza |
| Center | Abruzzo | 1,323 | Chieti; L'Aquila; Pescara; Teramo |
| Center | Lazio | 5,534 | Frosinone; Latina; Rieti; Rome; Viterbo |
| Center | Marche | 1,549 | Ancona; Ascoli Piceno; Fermo; Macerata; Pesaro-Urbino |
| Center | Tuscany | 3,675 | Arezzo; Florence; Grosseto; Livorno; Lucca; Massa Carrara; Pisa; Pistoia; Prato; Siena |
| Center | Umbria | 884 | Perugia; Terni |
| South & islands | Basilicata | 591 | Matera; Potenza |
| South & islands | Calabria | 2,006 | Catanzaro; Cosenza; Crotone; Reggio Calabria; Vibo Valentia |
| South & islands | Campania | 5,806 | Avellino; Benevento; Caserta; Naples; Salerno |
| South & islands | Molise | 321 | Campobasso; Isernia |
| South & islands | Puglia | 4,076 | Bari; Barletta-Andria-Trani; Brindisi; Foggia; Lecce; Taranto |
| South & islands | Sardinia | 1,665 | Cagliari; Carbonia-Iglesias; Medio-Campidano; Nuoro; Ogliastra; Olbia-Tempio; Oristano; Sassari |
| South & islands | Sicily | 5,029 | Agrigento; Caltanissetta; Catania; Enna; Messina; Palermo; Ragusa; Syracuse; Trapani |

*Table 2: List of the three leading subject categories for incidence in national SS, for each region (2006-2010 data)*

| Region | Subject category 1 | Subject category 2 | Subject category 3 |
| --- | --- | --- | --- |
| Abruzzo | Meteorology & atmospheric sciences (9.9%) | Dentistry, oral surgery & medicine (9.0%) | Neuroimaging (6.9%) |
| Basilicata | Engineering, petroleum (8.1%) | Entomology (6.7%) | Imaging science & photographic technology (6.0%) |
| Calabria | Engineering, manufacturing (10.4%) | Engineering, chemical (8.8%) | Engineering, industrial (6.7%) |
| Campania | Polymer science (20.5%) | Engineering, geological (16.0%) | Thermodynamics (15.7%) |
| Emilia Romagna | Orthopedics (33.3%) | Materials science, ceramics (32.7%) | Integrative & complementary medicine (26.7%) |
| Friuli Venezia Giulia | Engineering, marine (27.9%) | Physics, condensed matter (13.6%) | Physics, particles & fields (13.3%) |
| Lazio | Tropical medicine (37.5%) | Sport sciences (36.4%) | Parasitology (29.5%) |
| Liguria | Allergy (20.5%) | Robotics (15.2%) | Rheumatology (13.2%) |
| Lombardy | Ornithology (47.1%) | Medical informatics (34.9%) | Cell & tissue engineering (34.8%) |
| Marche | Fisheries (12.3%) | Medicine, legal (11.7%) | Geriatrics & gerontology (8.5%) |
| Molise | Engineering, petroleum (7.5%) | Forestry (4.7%) | Entomology (2.6%) |
| Piedmont | Materials science, textiles, paper & wood (36.0%) | Limnology (19.8%) | Mycology (19.4%) |
| Puglia | Computer science, cybernetics (15.4%) | Materials science, characterization & testing (14.2%) | Rehabilitation (13.3%) |
| Sardinia | Substance abuse (18.4%) | Agriculture, dairy & animal science (9.6%) | Mycology (6.5%) |
| Sicily | Engineering, marine (21.7%) | Ornithology (11.5%) | Chemistry, applied (11.3%) |
| Tuscany | Andrology (38.5%) | Robotics (29.0%) | Ergonomics (21.7%) |
| Trentino Alto Adige | Remote sensing (11.3%) | Computer science, cybernetics (11.0%) | Computer science, software engineering (8.4%) |
| Umbria | Engineering, petroleum (7.8%) | Mycology (6.2%) | Materials science, composites (5.2%) |
| Valle D'Aosta | Operations research & management science (0.3%) | Environmental studies (0.3%) | Biodiversity conservation (0.1%) |
| Veneto | Medical laboratory technology (26.9%) | Physics, nuclear (15.9%) | Limnology (15.7%) |

*Table 3: Top 20 subject categories by regional concentration of SS, with list of the three leading regions for national incidence*

| Subject category | Discipline | Region 1 | Region 2 | Region 3 | Cumul. | Gini |
|---|---|---|---|---|---|---|
| Ornithology | Biology | Lombardy (47.1%) | Emilia Romagna (19.3%) | Sicily (11.5%) | 77.9% | 0.83 |
| Andrology | Clinical Medicine | Tuscany (38.5%) | Emilia Romagna (20.5%) | Lazio (15.8%) | 74.8% | 0.73 |
| Orthopedics | Clinical Medicine | Emilia Romagna (33.3%) | Lazio (25.5%) | Lombardy (14.8%) | 73.6% | 0.71 |
| Engineering, Marine | Engineering | Friuli Venezia Giulia (27.9%) | Lombardy (23.4%) | Sicily (21.7%) | 73.0% | 0.78 |
| Tropical Medicine | Clinical Medicine | Lazio (37.5%) | Lombardy (22.3%) | Campania (10.5%) | 70.3% | 0.75 |
| Sport Sciences | Clinical Medicine | Lazio (36.4%) | Lombardy (17.4%) | Emilia Romagna (11.6%) | 65.5% | 0.70 |
| Integrative & Complementary Medicine | Clinical Medicine | Emilia Romagna (26.7%) | Lombardy (25.2%) | Veneto (13.3%) | 65.2% | 0.68 |
| Medical Laboratory Technology | Biomedical Research | Veneto (26.9%) | Lombardy (24.3%) | Tuscany (13.0%) | 64.3% | 0.66 |
| Robotics | Engineering | Tuscany (29.0%) | Lazio (19.7%) | Liguria (15.2%) | 63.9% | 0.71 |
| Cell & Tissue Engineering | Biology | Lombardy (34.8%) | Emilia Romagna (19.2%) | Veneto (9.0%) | 63.1% | 0.70 |
| Rehabilitation | Clinical Medicine | Lazio (24.3%) | Lombardy (22.8%) | Puglia (13.3%) | 60.5% | 0.65 |
| Materials Science, Textiles, Paper & Wood | Engineering | Piedmont (36.0%) | Lombardy (16.0%) | Emilia Romagna (7.6%) | 59.6% | 0.64 |
| Critical Care Medicine | Clinical Medicine | Lombardy (25.9%) | Piedmont (18.9%) | Veneto (14.8%) | 59.6% | 0.68 |
| Cardiac & Cardiovascular Systems | Clinical Medicine | Lombardy (32.3%) | Lazio (14.1%) | Emilia Romagna (12.5%) | 58.8% | 0.66 |
| Medical Informatics | Engineering | Lombardy (34.9%) | Tuscany (11.8%) | Lazio (11.2%) | 58.0% | 0.65 |
| Infectious Diseases | Biomedical Research | Lombardy (28.8%) | Lazio (21.2%) | Veneto (7.8%) | 57.8% | 0.61 |
| Anesthesiology | Clinical Medicine | Lombardy (30.0%) | Lazio (14.4%) | Tuscany (12.7%) | 57.0% | 0.65 |
| Health Care Sciences & Services | Clinical Medicine | Lombardy (25.9%) | Lazio (22.9%) | Emilia Romagna (8.1%) | 56.9% | 0.60 |
| Imaging Science & Photographic Technology | Physics | Lombardy (32.5%) | Lazio (13.4%) | Tuscany (10.9%) | 56.8% | 0.64 |
| Materials Science, Ceramics | Engineering | Emilia Romagna (32.7%) | Veneto (14.1%) | Piedmont (10.0%) | 56.7% | 0.62 |

*Table 4: Twenty subject categories with SS least concentrated at the level of regional distribution, with list of the three leading regions for national incidence*

| Subject category | Discipline | Region 1 | Region 2 | Region 3 | Cumul. | Gini |
|---|---|---|---|---|---|---|
| Marine & Freshwater Biology | Biology | Lombardy (12.2%) | Tuscany (10.1%) | Emilia Romagna (9.9%) | 32.2% | 0.44 |
| Entomology | Biology | Emilia Romagna (12.9%) | Piedmont (11.0%) | Veneto (10.1%) | 34.0% | 0.40 |
| Chemistry, Physical | Chemistry | Lombardy (13.0%) | Emilia Romagna (11.5%) | Tuscany (11.4%) | 35.8% | 0.45 |
| Engineering, Manufacturing | Engineering | Lombardy (14.9%) | Sicily (10.7%) | Calabria (10.4%) | 35.9% | 0.50 |
| Engineering, Electrical & Electronic | Engineering | Lombardy (14.5%) | Lazio (11.0%) | Emilia Romagna (10.4%) | 36.0% | 0.48 |
| Nanoscience & Nanotechnology | Engineering | Lombardy (14.3%) | Emilia Romagna (12.2%) | Friuli Venezia Giulia (10.4%) | 37.0% | 0.46 |
| Oceanography | Earth and Space Sciences | Emilia Romagna (17.2%) | Veneto (10.4%) | Friuli Venezia Giulia (9.5%) | 37.0% | 0.50 |
| Computer Science, Theory & Methods | Engineering | Tuscany (14.5%) | Lombardy (12.6%) | Emilia Romagna (9.9%) | 37.0% | 0.46 |
| Materials Science, Multidisciplinary | Engineering | Lombardy (15.0%) | Emilia Romagna (13.6%) | Piedmont (8.9%) | 37.5% | 0.45 |
| Water Resources | Earth and Space Sciences | Lombardy (14.0%) | Veneto (12.3%) | Emilia Romagna (11.3%) | 37.6% | 0.46 |
| Plant Sciences | Biology | Lazio (14.0%) | Tuscany (13.6%) | Lombardy (10.3%) | 37.8% | 0.47 |
| Engineering, Civil | Engineering | Lombardy (16.2%) | Emilia Romagna (11.2%) | Campania (10.5%) | 37.9% | 0.42 |
| Mechanics | Physics | Lazio (14.8%) | Emilia Romagna (12.2%) | Lombardy (11.3%) | 38.2% | 0.49 |
| Metallurgy & Metallurgical Engineering | Engineering | Lombardy (17.0%) | Liguria (11.6%) | Piedmont (9.8%) | 38.4% | 0.47 |
| Instruments & Instrumentation | Engineering | Lombardy (17.6%) | Emilia Romagna (10.6%) | Lazio (10.3%) | 38.5% | 0.49 |
| Agronomy | Biology | Emilia Romagna (15.0%) | Lombardy (12.8%) | Puglia (10.8%) | 38.5% | 0.47 |
| Thermodynamics | Physics | Campania (15.7%) | Piedmont (11.6%) | Veneto (11.4%) | 38.7% | 0.52 |
| Chemistry, Applied | Chemistry | Emilia Romagna (14.0%) | Lombardy (13.5%) | Sicily (11.3%) | 38.9% | 0.45 |
| Veterinary Sciences | Biology | Lombardy (15.5%) | Emilia Romagna (12.7%) | Veneto (10.9%) | 39.0% | 0.50 |
| Food Science & Technology | Biology | Emilia Romagna (17.1%) | Lombardy (12.2%) | Puglia (10.1%) | 39.4% | 0.45 |

*Table 5: Subject categories with SS most concentrated at the level of Provincial distribution, with list of the three leading provinces for national incidence (data 2006-2010)*

| Subject category | Discipline | Province 1 | Province 2 | Province 3 | Cumulative incidence | Gini |
|---|---|---|---|---|---|---|
| Engineering, Marine | Engineering | Trieste (26.2%) | Pavia (21.4%) | Messina (18.4%) | 66.0% | 0.95 |
| Orthopedics | Clinical Medicine | Bologna (28.9%) | Rome (24.5%) | Milan (9.7%) | 63.1% | 0.89 |
| Ornithology | Biology | Milan (33.4%) | Pavia (15.1%) | Palermo (10.4%) | 58.9% | 0.94 |
| Andrology | Clinical Medicine | Florence (25.5%) | Bologna (16.2%) | Rome (15.4%) | 57.1% | 0.91 |
| Sport Sciences | Clinical Medicine | Rome (35.1%) | Milan (12.7%) | Bologna (9.1%) | 57.0% | 0.87 |
| Tropical Medicine | Clinical Medicine | Rome (36.8%) | Naples (10.3%) | Pavia (9.5%) | 56.6% | 0.92 |
| Rehabilitation | Clinical Medicine | Rome (22.6%) | Milan (18.6%) | Bari (12.2%) | 53.4% | 0.85 |
| Robotics | Engineering | Pisa (21.6%) | Genoa (14.5%) | Rome (14.2%) | 50.3% | 0.90 |
| Engineering, Aerospace | Engineering | Pisa (18.1%) | Rome (16.9%) | Turin (14.8%) | 49.8% | 0.91 |
| Materials Science, Textiles, Paper & Wood | Engineering | Turin (28.6%) | Milan (14.5%) | Ravenna (6.4%) | 49.6% | 0.89 |
| Integrative & Complementary Medicine | Clinical Medicine | Bologna (19.6%) | Varese (17.9%) | Verona (10.4%) | 47.9% | 0.87 |
| Meteorology & Atmospheric Sciences | Earth and Space Sciences | Bologna (19.8%) | Varese (16.2%) | Rome (11.4%) | 47.4% | 0.88 |
| Developmental Biology | Biology | Milan (20.1%) | Rome (14.6%) | Naples (12.7%) | 47.4% | 0.88 |
| Parasitology | Clinical Medicine | Rome (27.4%) | Milan (9.9%) | Bari (9.7%) | 47.0% | 0.89 |
| Engineering, Petroleum | Engineering | Milan (20.7%) | Naples (12.9%) | Rome (12.5%) | 46.1% | 0.91 |
| Cell & Tissue Engineering | Biology | Milan (28.4%) | Modena (10.7%) | Rome (7.0%) | 46.1% | 0.87 |
| Substance Abuse | Clinical Medicine | Rome (25.5%) | Cagliari (13.4%) | Sassari (7.0%) | 45.9% | 0.88 |
| Medical Laboratory Technology | Biomedical Research | Milan (18.2%) | Padua (13.9%) | Verona (12.8%) | 44.9% | 0.85 |
| Infectious Diseases | Biomedical Research | Milan (20.2%) | Rome (19.5%) | Padua (4.6%) | 44.3% | 0.82 |
| Health Care Sciences & Services | Clinical Medicine | Rome (21.4%) | Milan (18.6%) | Bologna (4.0%) | 44.0% | 0.82 |

*Table 6: The first 20 Subject category-Region pairings for maximum level of difference of "SS per inhabitant" from the national average*

| Subject category | Region |
| --- | --- |
| Engineering, petroleum | Molise (1753.1%) |
| Engineering, marine | Friuli Venezia Giulia (1564.0%) |
| Forestry | Molise (1249.1%) |
| Allergy | Liguria (1220.8%) |
| Engineering, petroleum | Basilicata (1032.4%) |
| Astronomy & astrophysics | Friuli Venezia Giulia (991.4%) |
| Physics, particles & fields | Friuli Venezia Giulia (904.0%) |
| Physics, condensed matter | Friuli Venezia Giulia (884.8%) |
| Entomology | Basilicata (832.2%) |
| Andrology | Tuscany (831.6%) |
| Remote sensing | Trentino Alto Adige (828.5%) |
| Substance abuse | Sardinia (811.4%) |
| Rheumatology | Liguria (779.5%) |
| Imaging science & photographic technology | Basilicata (775.7%) |
| Engineering, geological | Basilicata (755.0%) |
| Computer science, cybernetics | Trentino Alto Adige (692.6%) |
| Physics, multidisciplinary | Friuli Venezia Giulia (682.8%) |
| Forestry | Basilicata (668.1%) |
| Engineering, petroleum | Umbria (665.8%) |
| Oceanography | Friuli Venezia Giulia (639.0%) |

*Table 7: The first 20 Subject category-Region pairings for maximum level of difference of "Output per inhabitant" from the national average*

| Subject category | Region |
| --- | --- |
| Forestry | Molise (1191.0%) |
| Engineering, marine | Liguria (1092.7%) |
| Allergy | Liguria (956.4%) |
| Engineering, petroleum | Molise (725.0%) |
| Materials science, textiles, paper & wood | Trentino Alto Adige (705.7%) |
| Astronomy & astrophysics | Friuli Venezia Giulia (684.5%) |
| Entomology | Molise (679.6%) |
| Physics, particles & fields | Friuli Venezia Giulia (677.4%) |
| Physics, multidisciplinary | Friuli Venezia Giulia (661.5%) |
| Oceanography | Friuli Venezia Giulia (657.0%) |
| Engineering, petroleum | Basilicata (655.2%) |
| Physics, condensed matter | Friuli Venezia Giulia (650.0%) |
| Medicine, legal | Molise (619.5%) |
| Dentistry, oral surgery & medicine | Abruzzo (587.6%) |
| Andrology | Tuscany (584.4%) |
| Substance abuse | Sardinia (575.9%) |
| Physics, mathematical | Friuli Venezia Giulia (548.5%) |
| Forestry | Trentino Alto Adige (542.6%) |
| Forestry | Basilicata (541.5%) |
| Rheumatology | Liguria (540.7%) |

*Table 8: The first 20 Subject category-Region pairings for maximum difference of SS/GDP from the national average*

| Subject category | Region |
| --- | --- |
| Engineering, petroleum | Molise (2267,9%) |
| Forestry | Molise (1616,0%) |
| Engineering, petroleum | Basilicata (1497,5%) |
| Engineering, marine | Friuli Venezia Giulia (1381,4%) |
| Entomology | Basilicata (1207,1%) |
| Allergy | Liguria (1159,9%) |
| Imaging science & photographic technology | Basilicata (1125,3%) |
| Engineering, geological | Basilicata (1095,2%) |
| Substance abuse | Sardinia (1075,1%) |
| Forestry | Basilicata (969,1%) |
| Astronomy & astrophysics | Friuli Venezia Giulia (875,7%) |
| Medicine, legal | Molise (799,9%) |
| Physics, particles & fields | Friuli Venezia Giulia (798,5%) |
| Entomology | Molise (788,4%) |
| Physics, condensed matter | Friuli Venezia Giulia (781,5%) |
| Andrology | Tuscany (765,4%) |
| Agronomy | Basilicata (763,2%) |
| Agriculture, dairy & animal science | Molise (753,4%) |
| Rheumatology | Liguria (740,6%) |
| Engineering, petroleum | Umbria (713,1%) |

*Table 9: The first 20 Subject category-Region pairings for maximum level of difference of "SS per inhabitant" from the national average*

| Subject category | Province |
|---|---|
| Engineering, Marine | Trieste (6592.3%) |
| Limnology | Verbania (3812.7%) |
| Physics, Condensed Matter | Trieste (3232.58%) |
| Physics, Particles & Fields | Trieste (3219.4%) |
| Robotics | Pisa (3169.0%) |
| Astronomy & Astrophysics | Trieste (2942.8%) |
| Materials Science, Ceramics | Ravenna (2671.3%) |
| Engineering, Aerospace | Pisa (2660.8%) |
| Physics, Multidisciplinary | Trieste (2471.6%) |
| Nanoscience & Nanotechnology | Trieste (2453.6%) |
| Engineering, Marine | Pavia (2408.2%) |
| Forestry | Viterbo (2363.2%) |
| Ergonomics | Pisa (2330.5%) |
| Oceanography | Trieste (2263.1%) |
| Horticulture | Viterbo (2226.6%) |
| Mining & Mineral Processing | Trieste (2197.6%) |
| Dentistry, Oral Surgery & Medicine | Siena (2179.98%) |
| Chemistry, Multidisciplinary | Trieste (2176.1%) |
| Andrology | Siena (2152.1%) |
| Environmental Studies | Siena (2081.6%) |